\documentclass[AMA,STIX2COL,Linenumbersoff,sort,compress]{MRM}
\articletype{Research Article}

\received{}
\revised{}
\accepted{}
\usepackage{color} 
\definecolor{black}{rgb}{0,0,0}
\definecolor{blue}{rgb}{0,0,0.8}
\definecolor{fuchsia}{rgb}{0.5,0,0.8}

\definecolor{red}{rgb}{0.8,0,0}

\definecolor{green}{rgb}{0,0.4,0}

\newcommand{\update}[1]{{\color{black}#1}}
\usepackage{times}
\usepackage{amssymb}
\usepackage{siunitx}
\usepackage{upgreek} 
\usepackage{ulem}
\usepackage{amsmath}
\usepackage{amsfonts}
\usepackage{braket}
\usepackage{fancyhdr}
\usepackage{multirow}
\usepackage{graphicx}

\DeclareMathOperator{\tr}{tr}
\newcommand{\n}{\hat{\mathbf{n}}}

\newcommand{\B}{\mathsf{B}}

\renewcommand{\L}{\mathsf{L}}
\renewcommand{\d}{\mathrm{d}}

\renewcommand{\r}{\mathbf{r}}
\newcommand{\q}{\mathbf{q}}

\newcommand{\lb}{\left[}
\newcommand{\rb}{\right]}
\newcommand{\lp}{\left(}
\newcommand{\rp}{\right)}
\newcommand{\N}{\mathsf{N}}
\newcommand{\Y}{\mathcal{Y}}
\newcommand{\g}{\hat{\mathbf{g}}}
\newcommand{\bz}{b^\circ}
\newcommand{\betaz}{\beta^\circ}
\newcommand{\Bz}{{\mathsf{B}^\circ}}
\newcommand{\qz}{\mathbf{q}^\circ}
\newcommand{\gz}{{\hat{\mathbf{g}}^\circ}}
\newcommand{\Gz}{{\mathbf{G}}^\circ}
\newcommand{\qqz}{{q}^\circ}
\newcommand{\GGz}{{{G}}^\circ}
\begin{document}


\title{What if each voxel were measured with a different diffusion protocol?}

\author[1]{Santiago Coelho}{\orcid{0000-0002-1039-4803}}
\author[1]{Gregory Lemberskiy}{\orcid{0000-0002-8336-7486}}
\author[2]{Ante Zhu}{\orcid{0000-0001-6251-2341}}
\author[3]{Hong-Hsi Lee}{\orcid{0000-0002-3663-6559}}
\author[2]{Nastaren Abad}{\orcid{0000-0002-7020-0492}}
\author[2]{Thomas K. F. Foo}{\orcid{0000-0002-9016-1005}}
\author[1]{Els Fieremans}{\orcid{0000-0002-1384-8591}}
\author[1]{Dmitry S. Novikov}{\orcid{0000-0002-4213-3050}}



\authormark{\textsc{COELHO et al.}}

\address[1]{\orgdiv{Center for Biomedical Imaging and Center for Advanced Imaging Innovation and Research (CAI$^2$R)}, \orgname{Department of Radiology, New York University School of Medicine}, \orgaddress{\state{New York}, \country{United States}}}
\address[2]{\orgdiv{GE HealthCare Technology and Innovation Center}, \orgname{Niskayuna}, \orgaddress{\state{New York}, \country{United States}}}
\address[3]{\orgdiv{Athinoula A. Martinos Center for Biomedical Imaging}, \orgname{Massachusetts General Hospital, Harvard Medical School}, \orgaddress{\state{Charlestown}, \country{United States}}}

\corres{Santiago Coelho, 660 First Avenue, New York, NY 10016, USA. \email{Santiago.Coelho@nyulangone.org}}


\finfo{This work was partially supported by \fundingAgency{National Institutes of Health, National Institute of Biomedical Imaging and Bioengineering} Grant/Award Number: \fundingNumber{P41-EB017183}, \fundingNumber{R01-EB027075}, and \fundingNumber{K99-EB036080}, and \fundingAgency{National Institute of Neurological Disorders and Stroke} Grant/Award Number: \fundingNumber{R01-NS088040}, and \fundingAgency{Office of the Director and National Institute of Dental \& Craniofacial Research} Grant/Award Number: \fundingNumber{DP5-OD031854}.}


\abstract[Abstract]{
	{\normalsize \section{Purpose} 
Expansion of diffusion MRI (dMRI) both into the realm of strong gradients, and into accessible imaging with portable low-field devices, brings about the challenge of gradient nonlinearities. Spatial variations of the diffusion gradients make diffusion weightings and directions non-uniform across the field of view, and deform perfect shells in the $q$-space designed for isotropic directional coverage. 
Such imperfections hinder parameter estimation: Anisotropic shells hamper the deconvolution of  fiber orientation distribution function (fODF), while brute-force retraining of a nonlinear regressor for each unique set of directions and diffusion weightings is computationally inefficient.
\section{Methods} 
Here we propose a protocol-independent parameter estimation (PIPE) method that enables fast parameter estimation for the most general case where the scan in each voxel is acquired with a different protocol in $q$-space. PIPE applies for any spherical convolution-based dMRI model, irrespective of its complexity, which makes it suitable both for white and gray matter in the brain or spinal cord, and for other tissues where fiber bundles have the same  properties within a voxel (fiber response), but are distributed with an arbitrary fODF. 
We also derive a parsimonious representation that isolates isotropic and anisotropic effects of gradient nonlinearities on multidimensional diffusion encodings.
\section{Results} \update{Applied to {\it in vivo} human MRI with linear tensor encoding} on a high-performance gradient system, PIPE maps fiber response and fODF parameters for the whole brain in the presence of significant gradient nonlinearities in under $3$ minutes. 
\section{Conclusions} PIPE enables fast parameter estimation in the presence of arbitrary gradient nonlinearities, eliminating the need to arrange dMRI in shells or to retrain the estimator for different protocols in each voxel. PIPE applies for any model based on a convolution of a voxel-wise fiber response and fODF, and data from varying \update{$b$-values}, diffusion/echo times, and other scan parameters.}}
\maketitle
\footnotetext{\textbf{Abbreviations:}~\hbox{dMRI, diffusion Magnetic Resonance Imaging; SM,Standard Model; SVD; singular value decomposition; FOV, field of view}}
\update{This manuscript has approximately 5000 words.}

\clearpage
\newpage

\section{Introduction}\label{sec:Intro}
\noindent
The ability to non-invasively probe the random motion of water molecules \update{within} tissues makes diffusion MRI (dMRI) sensitive to tissue micro-architecture\cite{KISELEV2017,NOVIKOV2019,ALEXANDER2019,NOVIKOV2021,KISELEV2021,WEISKOPF2021}. 
Typical experimental settings detect such motion at a scale of $1-50$ micrometers, making dMRI sensitive, and possibly specific, to disease processes originating at this scale, and thereby provide biomarkers of pathological processes \cite{ASSAF2008b,JONES2010,JELESCU2017,NILSSON2018,NOVIKOV2018b}. Biophysical models of diffusion have the potential to deliver the specificity that would aid in early diagnosis \cite{JELESCU2017}. This prompts the development of dMRI acquisitions and models that strive not only for sensitivity but also for specificity\cite{NOVIKOV2018b}.


The development of ultra-high gradients for tissue microstructure imaging\cite{FOO2020,HUANG2021,JIA2021,FEINBERG2023,RAMOS2025}, and the availability of cost-effective, portable scanners\cite{SARRACANIE2015,COOLEY2021,LIU2021,ZHAO2024}, has created unprecedented opportunities for dMRI\cite{JONES2018}. At the high-performance end, {\it in vivo} dMRI exhibits an improved resolution\cite{MCNAB2013}, as well as the sensitivity to axon diameters\cite{VERAART2020}, diffusion time-dependence \cite{DAI2023}, water exchange\cite{CHAN2025}, and the magnetization localized near cell walls \cite{LEE2022}. \update{At the opposite end of the spectrum, ultra-low-field systems\cite{CAMPBELLWASHBURN2023}, gaining traction due to their affordability and accessibility, offer the possibility of diffusion tensor imaging and fiber tracking\cite{ABATE2024,GHOLAM2025}.}


\update{Head-only MRI systems offer the opportunity to optimize gradient performance at both high-performance and low field ends.} 
These gradients are designed to have good linearity in about $25\,$cm-diameter sphere, compared to $40\!-\!50\,$cm for whole-body gradients \cite{MCNAB2013}. Therefore, in the peripheral parts of the brain, {\it gradient nonlinearities} are non-negligible\cite{JONES2018}, (Fig.\ref{fig:gradientNonlinearities}). Away from the bore isocenter, applied diffusion weightings deviate from the nominal settings; this deviation further depends on the gradient direction\cite{BAMMER2003}. 
Thus, actual diffusion weightings and measurement protocols can vary significantly across the field of view (FOV), and may even not consist of conventional shells in the diffusion $q$-space designed for isotropic directional sampling. 
The anisotropy of shells, and the spatially varying diffusion weightings and directions, make diffusion processing\cite{RUDRAPATNA2021} and parameter estimation \update{of nonlinear models} challenging.

\update{As the logarithm of the DTI signal linearly depends on the diffusion encoding, estimated diffusion tensors can be corrected for gradient nonlinearities {\it a posteriori} in each voxel\cite{BAMMER2003}. 
For essentially nonlinear multi-compartment models, however, post-hoc parameter corrections are challenging, and the entire estimation process must be reconsidered. 
In this work, we focus on the} 
overarching \emph{spherical convolution} framework for modeling diffusion in brain tissue\cite{TOURNIER2004,ANDERSON2005,TOURNIER2007,JEURISSEN2014,NOVIKOV2019}, which assumes that each voxel contains a collection of identical fiber bundles with arbitrary orientations described by the fiber Orientation Distribution Function (fODF), Fig.~\ref{fig:convFramework}. The dMRI signal can be expressed as a convolution over the unit sphere $\mathbb{S}^2$ of the fiber bundle response function (the \textit{kernel}) and the fODF. The generality of this framework allows arbitrary number of gaussian or non-gaussian diffusion compartments, possibility of water exchange, diffusion time-, inversion time- or echo time-dependence, etc. This has enabled its application to both white matter (WM)\cite{JESPERSEN2007,FIEREMANS2011,ZHANG2012,KADEN2016,JENSEN2016,REISERT2017,NOVIKOV2018,VERAART2017} and gray matter (GM)\cite{PALOMBO2020,JELESCU2022,OLESEN2022}.

Conventional fiber responses depend nonlinearly on microstructural parameters, making it difficult to estimate these parameters robustly using standard maximum likelihood approaches. Additionally, many of these regression-based techniques assume that the data is acquired in spherical shells in the $q$-space, to ensure isotropic  coverage and a robust fODF deconvolution. Gradient nonlinearities can distort these shells, causing the actual diffusion weighting to vary with the direction and the spatial location across the imaging volume. At typical signal-to-noise ratios (SNR), data-driven regression methods have been shown to achieve lower mean-squared error (MSE) than likelihood-based methods\cite{COELHO2021a,LIAO2024}. However, retraining such machine learning (ML)-based estimators independently for each voxel to accommodate protocol variations would result in a prohibitively high computational cost.

Here we propose a {\it protocol-independent parameter estimation} (PIPE) ML framework that enables fast parameter estimation of convolution-based biophysical models, for which every voxel can be acquired with a different dMRI protocol, thereby enabling applications on MRI systems with arbitrary gradient nonlinearities\cite{COELHO2023b}. PIPE applies for any biophysical model where fiber bundles in a given voxel have the same properties, but are distributed with an arbitrary fODF. Both the fiber response and fODF can vary between voxels, thereby ensuring the applicability of PIPE to practically all white matter and gray matter dMRI models\cite{JESPERSEN2007,FIEREMANS2011,ZHANG2012,KADEN2016,JENSEN2016,REISERT2017,NOVIKOV2018,VERAART2017,PALOMBO2020,JELESCU2022,OLESEN2022}.

PIPE relies on the singular value decomposition (SVD) based separation between the protocol and model parameters in any model expression\cite{COELHO2023}, such that the training is done only once on the model-parameter part, while the protocol-dependent part is allowed to vary from voxel to voxel. Furthermore, the data is not constrained to be acquired in any fashion (such as shells), and there are no limits on the gradient nonlinearities as long as the protocol is well defined for each voxel. PIPE  incorporates varying diffusion times, echo times, and other scan parameters within the protocol,  \update{and is  naturally suited for  the most widespread linear tensor encoding (attained, e.g., via PGSE\cite{STEJSKAL&TANNER1965}), being generalizable to any axially symmetric multidimensional diffusion encodings}. 

In what follows, we will describe the method, evaluate its accuracy and precision via synthetic noise propagation, and apply it to {\it in vivo} human brain \update{PGSE-dMRI} acquired on a high-performance  system with notable gradient nonlinearities.

\section{Methods}\label{sec:Methods}
\subsection{Machine Re-Learning?}
\noindent
Why should one worry about having different sets of diffusion directions and $b$-values in every voxel? Theoretically, one can use an unbiased estimator for each voxel separately; after all, a well-validated model is the best way to factor out the differences in protocol\cite{NOVIKOV2018b}. 
Conventional maximum likelihood estimators\cite{KAY1993} are asymptotically unbiased and can be employed. Given the dMRI noise model, maximizing the likelihood across all measurements defines a mapping from the space of noisy measurements to model parameters. 

While statistically rigorous, max-likelihood estimation suffers from a few drawbacks. First, finding a global maximum for hundreds of measurements and at least a few model parameters is computationally intense for strongly nonlinear dMRI models, and must be performed in each voxel separately. Second, to speed up the search for the global likelihood maximum, one has to initialize the search within the right ``basin of attraction" around the ground truth, which is {\it a priori} unknown; this can make the outcome dependent on the initialization. Third, even with the appropriate initialization, the strongly nonlinear nature of diffusion models and a relatively low SNR make the estimated parameters veer off quite far from the ground truth along the ``shallow directions" in the likelihood landscape\cite{JELESCU2016a}, often into unphysical domains; this is typically cured by introducing priors or regularization terms  into the objective function \cite{MOZUMDER2019}, making the outcome dependent on their assumptions.

Alternatively, data-driven supervised ML regressions ``learn" the mapping from noisy measurements to model parameters. The mapping is provided by a suﬃciently flexible regressor, such as a neural network or a high-order multivariate polynomial, based on the  training data generated with the forward signal and noise models of interest. The regression coefficients are then chosen such that mean-squared error (MSE) or other error metrics are minimized over the training data \cite{GOLKOV2016,REISERT2017}. While the training can be slow and computationally intense, the estimation is typically fast, with the learned regression evaluated once per voxel. 

For realistic dMRI voxel-wise SNR ($\sim25-100$ for unweighted images), ML regressions outperform conventional  estimation approaches, even though the estimated parameters may be biased by the training set at low SNR\cite{COELHO2021a}. Multiple research groups have applied variations of these regressions to various biophysical models aimed at data from clinical scanners \cite{REISERT2017,PALOMBO2020,DEALMEIDA2021b,GYORI2022,COELHO2022}. However, applying ML regressions in the case of significant gradient nonlinearities is nontrivial: each voxel within the field of view may end up with a distinct set of diffusion weightings and directions. Thus, the dimensionality of the mapping one needs to learn increases dramatically:  e.g., in a FOV of $100\times100\times100$ voxels one ends up with 1 million unique protocols. Even when the gradient nonlinearities are slowly varying, one would have to ``re-learn" the mapping for thousands of distinct protocols for local patches with sufficiently different sets of gradient directions and weightings. 
Naively training an ML estimator for each voxel would take too long to compute. The PIPE framework addresses this challenge.

\begin{figure*}[htbp]
	\includegraphics[width=\linewidth]{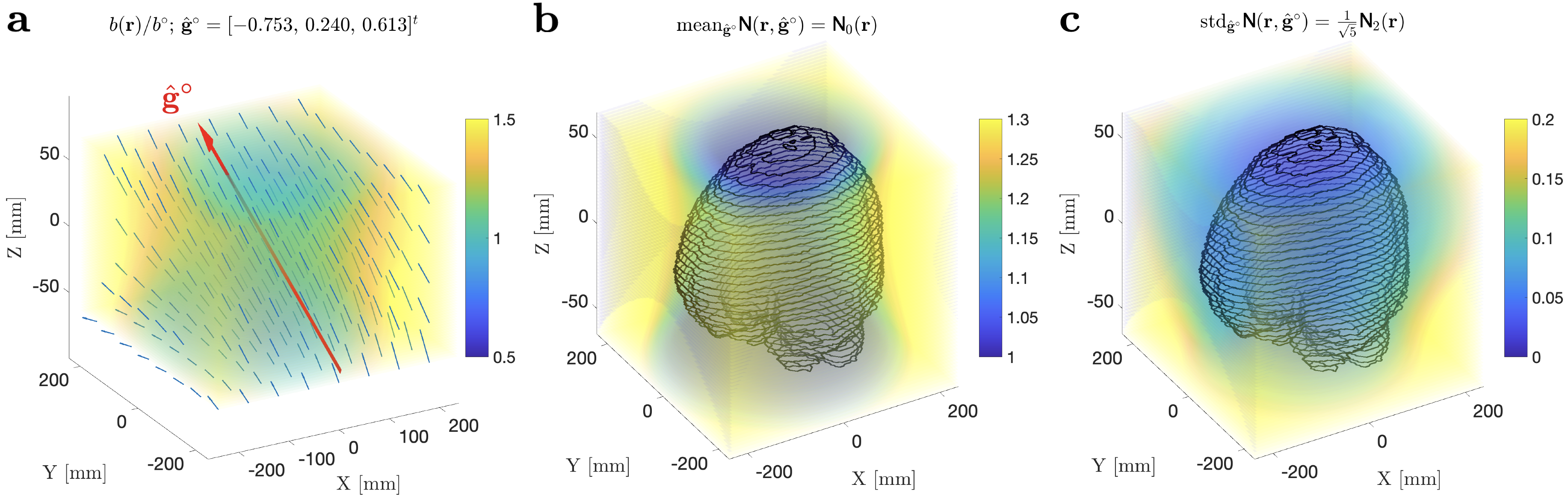}
	\caption[caption FIG 3]{
	Deformation of LTE $b$-value and direction by the gradient nonlinearities across the FOV \update{based on the measured $\mathsf{L}_{ij}(\r)$} for a high-performance head-only system\cite{FOO2020}. 
	LTE  weighting $b(\r,\gz)$, Eq.~(\ref{bLTE=}), varies differently across the FOV depending on its nominal direction $\gz$. 
	(a) An example of the deformation $\L(\r) \gz$ of a generic direction $\gz$, and the corresponding dimensionless ratio $b(\r,\gz)/\bz$ (color). 
	(b) The  mean over all possible LTE directions $\gz\in\mathbb{S}^2$ of the  ratio $b(\r,\gz)/\bz$,  Eq.~(\ref{bLTE=}), represented by the isotropic component $\N_0$ of $\N(\r)$, Eq.~(\ref{Nrotinvs0}). The brain contour of the volunteer is drawn for reference. 
	(c) Relative directional variations of the LTE $b$-value (\ref{bLTE=}), characterized by the standard deviation of $\N(\r,\gz)$ over all possible LTE directions $\gz$ (square root of the variance (\ref{Nrotinvs})). This quantity is determined by the anisotropic part $\N^{(2)}(\r)$ of $\N(\r)$ and is proportional to the invariant $\N_2(\r)$, Eq.~(\ref{Nrotinvs2}). 
}
	\label{fig:gradientNonlinearities}
\end{figure*}


\subsection{Gradient nonlinearities }\label{ss:GNC}
\noindent
Nonuniform magnetic field gradients introduce artifacts in dMRI. This causes not only image distortions\cite{THILAKA1994} but also spatially varying errors in the direction and strength of the applied diffusion encoding\cite{BAMMER2003}. Thus, the actual gradient ${\bf G}(\r)=\nabla B_z(\r)$ affecting the local Larmor frequency (and responsible for the diffusion encoding) differs both in its strength and orientation from the nominal gradient $\Gz$. Since the Maxwell's equations are linear, the deviations from the nominal $\Gz$ can be parametrized by the gradient coil tensor field\cite{BAMMER2003} $\mathsf{L}_{ij}(\mathbf{r})$, which relates the actual to the nominal gradient in the coordinate system of the gradient coils:
\begin{equation}\label{eq:Gactual}
	G_i(\r) =  \mathsf{L}_{ij}(\r) \, \GGz_j \, ,
\end{equation}
where the summation over repeated indices is implied, and 
\begin{equation} \label{def_Lij}
	\mathsf{L}_{ij}(\r) \equiv {\partial G_i(\r) \over \partial \GGz_j} = {\partial B_z(\r) \over \partial x_i} \cdot \frac1{\GGz_j} \,. 
\end{equation}
Introducing the wave vector $\mathbf{q}(t)$ as the antiderivative, 
\[
q_i(t) \equiv \int_0^t\! \d \tau\, \gamma G_i(\tau) = \mathsf{L}_{ij} \, \qqz_j(t) , \ \  \qqz_j = \int_0^t\! \d \tau\, \gamma \GGz_j(\tau) ,
\] 
where $\gamma$ is gyromagnetic ratio, one obtains the diffusion-encoding $\mathsf{B}$-tensor\cite{MITRA1995,MORI1995,WONG1995,CHENG1999} as a spatially varying field\cite{BAMMER2003}
\begin{equation}\label{Bactual}
	\mathsf{B}(\r)  \equiv \int\! \d t\, \q(t)\otimes \q(t) =  \mathsf{L}(\r) \, \Bz \, \mathsf{L}^t(\r) 
\end{equation}
in terms of the nominal $\mathsf{B}$-tensor\cite{WESTIN2016,TOPGAARD2017}
\begin{equation} \label{Bz}
	\Bz  = \int\! \d t\,  \qz(t)\otimes \qz(t) \,. 
\end{equation}

The rank of $\B$ reflects how many dimensions of the diffusion process are being simultaneously probed; $\mathrm{rank}\,\B>1$ means probing the diffusion along more than one dimension. 
Let us consider \emph{linear tensor encoding} (LTE, $\mathrm{rank}\,\B=1$), for which pulsed gradient spin echo (PGSE)\cite{STEJSKAL&TANNER1965} is the most common gradient waveform. The nominal LTE tensor $\Bz = \bz \, \gz \otimes {\gz}$, defined by its nominal unit direction $|\gz|=1$ and  $b$-value $\bz$, becomes
\begin{equation}
	\begin{aligned}\label{BLTE}
		\mathsf{B}(\r) &=  \bz \,  \L(\r)\big(\gz \otimes {\gz} \big) \L^t(\r) \,.
	\end{aligned}
\end{equation}
Remarkably, the LTE rank remains unchanged by the gradient nonlinearities: $\mathrm{rank}\,\B(\r)=1$ for any $\L(\r)$, when $\mathrm{rank}\,\Bz=1$.   
Both the LTE direction $\L \gz$ and the $b$-value 
\begin{equation}\label{bLTE}
\begin{aligned}
		b\big(\r,\gz\big)  = \mathrm{tr}\, \B(\r) &=\bz \, \g^{\circ\,t } \, \N(\r) \, \gz \,, \\
		\N(\r) &= \L^t(\r) \L(\r) \,, 
\end{aligned}
\end{equation}
get modified by $\L(\r)$. Here, we introduced the  \textit{nonlinearity tensor field} $\N_{ij}(\r)=\L_{ki}(\r) \L_{kj}(\r)$, which at every point $\r$ is a symmetric $3\times3$ matrix, determined by 6 independent parameters, instead of 9 parameters in $\L_{ij}$. 
It admits the irreducible decomposition (we follow the notations of Ref.~\cite{COELHO2025}):
\begin{equation}
	\label{Ndecomp}
		\N_{ij} = \N^{(0)}_{ij} +  \N^{(2)}_{ij} \equiv \N_0  \delta_{ij}+ \sum_{m=-2}^2 \N^{2m} \Y^{2m}_{ij} \,. 
\end{equation}
Decomposition (\ref{Ndecomp}) separates the isotropic part $\N^{(0)}$, defined by a single parameter $\N_0=\frac13 \tr \N$, 
from its anisotropic counterpart $\N^{(2)}$, defined by five $\N^{2m}$. 
Here, $\delta_{ij}$ is Kronecker's delta, and $\Y^{2 m}_{ij}$ are the symmetric trace-free basis tensors \cite{THORNE1980,COELHO2025}, such that 
$\Y^{2 m}_{ij} \hat g^\circ_i \hat g^\circ_j = Y_{2 m}(\gz)$ for a unit vector $\gz \in \mathbb{S}^2$ yield complex spherical harmonics (SH) $Y_{2 m}(\gz)$. 
We employ Racah normalization for SH and for the basis tensors $\Y^{\ell m}_{i_1\dots i_\ell}$, as in Ref.~\cite{COELHO2025}: 
\begin{equation} \label{racah}
\int_{\mathbb{S}^2} \d\n \, Y_{\ell m}^*(\n) Y_{\ell' m'}(\n) = {\delta_{\ell \ell'} \, \delta_{m m'} \over 2\ell +1} \,, \quad \d\n = {\d\Omega \over 4\pi} \,, 
\end{equation}
where $\d\Omega = \sin\theta \d\theta \d\phi$ is the standard measure on the unit sphere $\mathbb{S}^2$, such that $\int_{\mathbb{S}^2} \d\n \equiv 1$, 
and $Y_{00}(\n)\equiv 1$. 
As a result, the LTE $b$-value (\ref{bLTE}) can be recast as 
\begin{equation}\label{bLTE=}
\begin{aligned}
		b(\r,\gz) =\bz\, \N(\r, \gz)\,, \quad \N &=  \N^{(0)}(\r)  + \N^{(2)}(\r,\gz) \,, \\ 
		\N^{(2)}(\r, \gz) =& \sum_{m=-2}^2 \N^{2m}(\r) Y_{2m}(\gz) \,. 
\end{aligned}
\end{equation}
Namely, at each point $\r$ in the FOV, the LTE $b$-value depends on the nominal direction $\gz$ via an ellipsoid  
$\N(\r,\gz)=\N_{ij}(\r) \hat g^\circ_i \hat g^\circ_j$, 
such that its  average over all  $\gz\in \mathbb{S}^2$ is 
\begin{equation}  \label{Nrotinvs0}
		\mathrm{mean}_{\gz}\,\N(\gz) 
		= \int_{\mathbb{S}^2} \d \gz\,  \N(\gz) 
		= \N_{0} \,, 
\end{equation}
and the directional variance 
\begin{equation} \label{Nrotinvs}
\mathrm{var}_{\gz}\,\N(\gz) =  \int_{\mathbb{S}^2} \d \gz\,  \big(\N(\gz) - \N_0\big)^2 = \tfrac15 {\N_2}^2 \, ,
\end{equation}
where \update{we dropped the $\r$-dependence for brevity, and }
\begin{equation}\label{Nrotinvs2}
	{\N_{2}}^2 = \left\| \N^{2m} \right\|^2= \sum_{m=-2}^2 \N^{2m*} \, \N^{2m}  
\end{equation}
is the $\ell=2$  invariant of tensor $\N$, cf. Ref.\cite{COELHO2025}. 
Figure \ref{fig:gradientNonlinearities}a shows the dimensionless ratio $b(\r,\gz)/\bz=\N(\r,\gz)$ for a generic $\gz$, as well as the vector field $\L(\r)\, \gz$ distorted relative to $\gz$, 
for the system\cite{FOO2020} used in this study. 
Figure~\ref{fig:gradientNonlinearities}b,c shows  maps of mean and standard deviation of 
$b(\r,\gz)/\bz$, Eqs.~(\ref{bLTE=})--(\ref{Nrotinvs2}).

Throughout the main text, we limit ourselves to LTE, as the shape of the $\B$-tensor (\ref{BLTE}) is determined by a single unit direction 
\begin{equation}\label{g}
\g(\r) = {\L(\r) \gz \over  |  \L(\r) \gz |} \,, 
\end{equation} 
and remains LTE (rank-1) for arbitrary $\L(\r)$. In Appendix \ref{AppBactual}, we outline how the $b$-value and shape of an arbitrary $\Bz$ tensor are affected by $\L(\r)$.  
As one may expect, a generic $\L(\r)$ removes any special symmetries of the nominal $\Bz$ tensor. Hence, the moment we use beyond-LTE encodings (\ref{Bz}) with $\mathrm{rank}\,\Bz>1$, we should assume that the actual $\B$-tensor (\ref{Bactual}) is neither axially- nor spherically-symmetric; working with generic-shape $\B$ tensors would take us away from a familiar spherical convolution, as described below. 

\update{The gradient coil tensor field $\L_{ij}(\r)$ is obtained in a  calibration experiment, which needs to be performed only once. 
A typical approach is to measure the static field produced by each gradient coil, represent these fields via harmonic polynomials\cite{ROMEO1984}
$r^{\ell} Y^{\ell m}(\hat\r)$ 
(as the solutions of Laplace equation), and compute the derivatives (\ref{def_Lij}) to get the $\r$-dependent $\L_{ij}(\r)$.}

\subsection{Spherical convolution: kernel -- fODF factorization}
\begin{figure}[t!!]
	\centering
	\includegraphics[width=\linewidth]{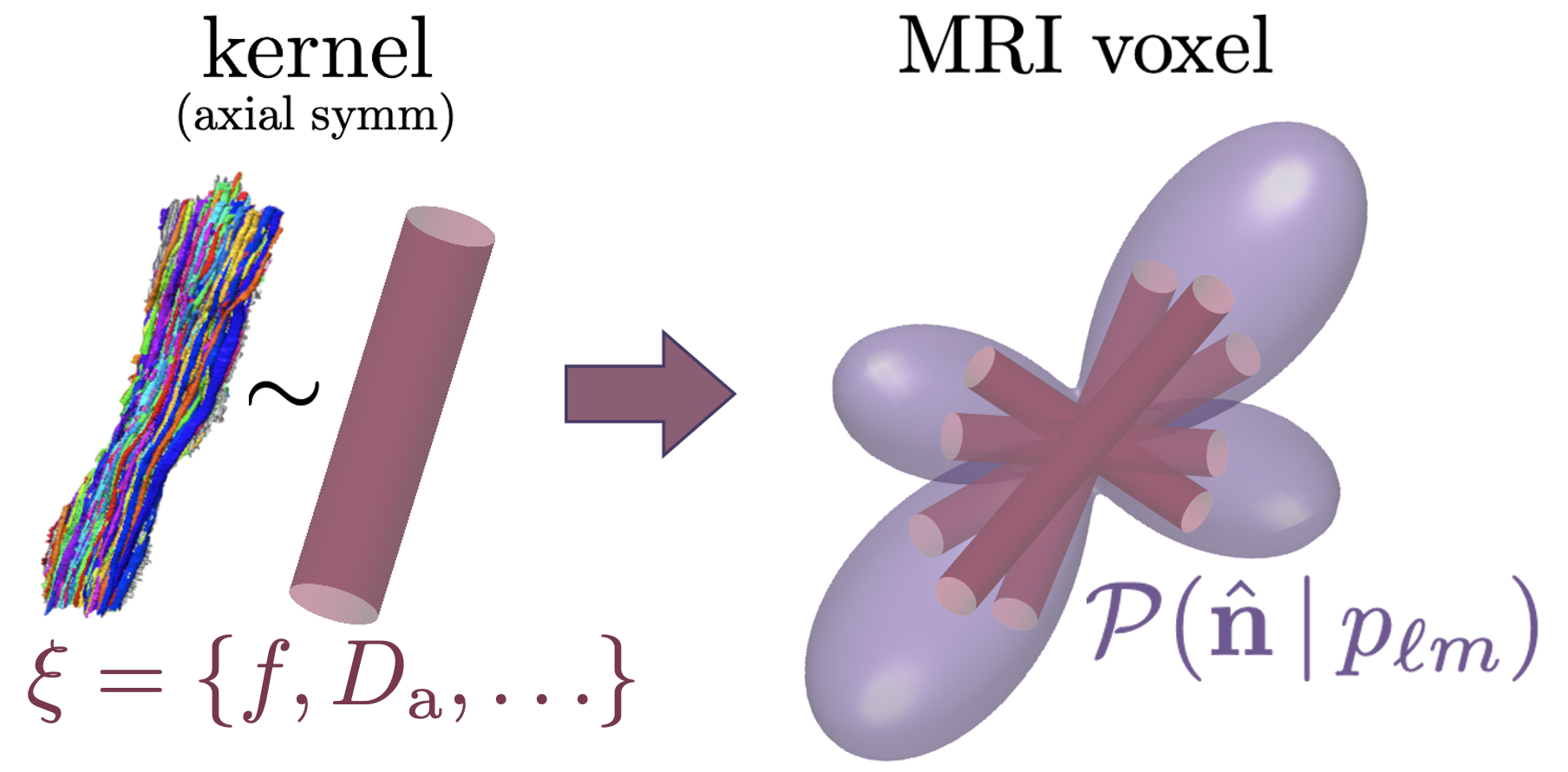}
	\caption[caption FIG 0]{PIPE central assumption: dMRI signal is  a spherical convolution of an axially-symmetric fiber response (kernel $\mathcal{K}$), characterized by scalar model parameters $\xi$, and an arbitrary fODF ${\cal P}(\n)$ characterized by its spherical harmonics coefficients $p_{\ell m}$. 
	}%
	\label{fig:convFramework}
\end{figure}

\noindent
PIPE applies for any microstructure model that can be represented as a spherical convolution. This functional form, 
empirically introduced at the level of signal representations \cite{TOURNIER2004,ANDERSON2005,TOURNIER2007,JEURISSEN2014}, 
encompasses the vast majority of biophysical models for white matter \cite{JESPERSEN2007,FIEREMANS2011,ZHANG2012,KADEN2016,JENSEN2016,REISERT2017,NOVIKOV2018,VERAART2017} and gray matter \cite{PALOMBO2020,JELESCU2022,OLESEN2022}.  
Namely, a voxel is assumed to contain a collection of identical fiber bundles with arbitrary fODF (Fig.~\ref{fig:convFramework}). 
Under the assumption of every fiber bundle in a voxel having the same microstructure (justified in Ref.~\cite{CHRISTIAENS2020}), one introduces the fiber response kernel $\mathcal{K}$ that depends on the \emph{mutual orientation} of the fiber direction ${\bf \hat n}$ and  the direction ${\bf \hat g}$ of the diffusion encoding. 
Because this response does not depend on  ${\bf \hat n}$ and ${\bf \hat g}$ separately, but only on their relative ``distance" 
$\hat{\mathbf{g}}\cdot\hat{\mathbf{n}}$ on a unit sphere, we can represent 
the dMRI signal as a convolution over the unit sphere $\mathbb{S}^2$ of the kernel with the fODF, Fig. \ref{fig:convFramework}: 
\begin{equation}\label{eq:DWIasConvolution}
	S(b,\hat{\mathbf{g}}\mid\,\!\xi,p_{\ell\!\,m})=\int_{\mathbb{S}^2}\!\mathrm{d}\hat{\mathbf{n}}\,\mathcal{K}(b,\hat{\mathbf{g}}\cdot\hat{\mathbf{n}}\mid\,\!\xi)\,\mathcal{P}(\hat{\mathbf{n}}\mid\,\!p_{\ell\,\!m}) \,.
\end{equation}
Here $b$ and $\hat{\mathbf{g}}$, Eqs.~(\ref{bLTE=}) and (\ref{g}), define the measurement, 
$\xi$ are the scalar \emph{tissue parameters} describing the kernel $\mathcal{K}$ (e.g., compartment fractions and diffusivities), 
and 
\begin{equation}\label{fODF}
\mathcal{P}(\hat{\mathbf{n}}\mid\,\!p_{\ell\,\!m})=\sum_{\ell=0,2,4,\dots} \sum_{m=-\ell}^\ell p_{\ell m}\,Y_{\ell\,\!m}(\hat{\mathbf{n}})
\end{equation} 
is the fODF  parametrized using coefficients $p_{\ell\,\!m}$ 
in the 
SH basis $Y_{\ell\,\!m}(\hat{\mathbf{n}})$, Eq.~(\ref{racah}), 
with $p_{00}\equiv 1$ (fODF normalization).

In Eq.~(\ref{eq:DWIasConvolution})  we can further include the diffusion time $t$, echo time $T_E$ and inversion time $T_I$ dependence of the signal. 
From now on, for brevity we will write $\mathcal{K}(b, \hat{\mathbf{g}}\cdot\hat{\mathbf{n}} \mid \xi)$, 
implying that $b$ stands for all scalar \emph{protocol parameters}, $b\to \{b, T_E, T_I, t, \dots \}$ defining the measurement, 
such that, e.g., compartments within the fiber bundle can have different $T_2$ and $T_1$ values\cite{VERAART2017,LAMPINEN2020,COELHO2024} (with the respective compartment relaxation times included as tissue parameters $\xi$). 
The dependence on PGSE diffusion time $t$ and pulse duration $\delta$ allows one to incorporate different classes of structural disorder in different compartments \cite{NOVIKOV2014,LEE2018,LEE2020,LEE2020b,CHAN2024},  as well as exchange between them \cite{JELESCU2022,OLESEN2022,NOVIKOV2023,LEE2025}, with the corresponding model parameters added to $\xi$. 

Much like convolutions become products under the ordinary Fourier transform, the convolution (\ref{eq:DWIasConvolution}) becomes a  product\cite{TOURNIER2004,ANDERSON2005,REISERT2017,NOVIKOV2018} in the SH basis: 
\begin{equation}\label{eq:SHfactorization}
		S_{\ell\,\!m}(b \mid \xi,p_{\ell m}) = \mathcal{K}_{\ell}(b,\mid \xi)\,p_{\ell\,\!m} \,, 
\end{equation}
where $S_{\ell m}$ are the signal's SH coefficients 
\begin{equation}\label{eq:SHfactorization2}
		S\left(b,\hat{\mathbf{g}}\mid\,\!\xi,p_{\ell\!\,m}\right) = \sum_{\ell\,\!m}S_{\ell\,\!m}(b \mid \xi,p_{\ell m})\,Y_{\ell\,\!m}(\hat{\mathbf{g}}) 
\end{equation}
with respect to the  directions (\ref{g}), and $\mathcal{K}_{\ell}\equiv \mathcal{K}_{\ell 0}$ are the projections of the kernel (aligned with the ${\bf \hat z}$ axis) onto Legendre polynomials $P_\ell(\zeta)=Y_{\ell 0}(\zeta)$, where $\zeta = \hat{\mathbf{g}}\cdot {\bf \hat z}$: 
\begin{equation}\label{eq:KrotInvs}
		\mathcal{K}_{\ell}(b \mid \xi) = \int_0^1 \,\d \zeta \, \mathcal{K}(b, \zeta \mid \xi) \, P_\ell(\zeta)  \, ,
\end{equation}
such that the kernel is expanded as 
\begin{equation}\label{eq:Kexpansion}
\mathcal{K}(b, \hat{\mathbf{g}}\cdot\hat{\mathbf{n}} \mid \xi) 
= \sum_{\ell = 0,2,4,\hdots} 	\mathcal{K}_{\ell}(b \mid \xi)\, P_\ell(\hat{\mathbf{g}}\cdot\hat{\mathbf{n}}) \, .
\end{equation}
The factorization (\ref{eq:SHfactorization}), essential for what follows, 
is valid as long as the kernel possesses axial symmetry ($\mathcal{K}_{\ell m}=0$ for nonzero $m$ in the fiber basis $\n={\bf \hat z}$). 
Note that beyond this 
assumption, the microstructure of the  fiber bundle is not constrained --- it can have an arbitrary number of compartments (with gaussian or non-gaussian diffusion), they can be coupled by exchange, have distinct relaxation properties, etc. 

We note in passing that the convolution for generic $\B$-tensors and non-axially symmetric kernels has to be defined on the SO(3) rotation group manifold 
$\mathbb{S}^3/\mathbb{Z}_2$ (the 3-dimensional sphere $\mathbb{S}^3$ with the antipodal points identified, $\mathbb{Z}_2=\{1,\ -1\}$), 
rather than on the 2-dimensional sphere $\mathbb{S}^2$. The corresponding Fourier transform involves a product of non-commuting matrices (the coefficients in the Wigner functions' basis), as the SO(3) group is non-abelian. 
For  axially symmetric response kernels, invariant under the SO(2) rotation around the fiber bundle axis, 
the convolution becomes effectively defined\cite{HEALY1998} on a quotient space $\mathbb{S}^2\cong\mathrm{SO(3)/SO(2)}$ that is equivalent to the 2-dimensional sphere  $\mathbb{S}^2$, which explains the integration over fODF directions in Eq.~(\ref{eq:DWIasConvolution}).

\begin{figure*}[htbp]
	\centering
	\includegraphics[width=0.9\textwidth]{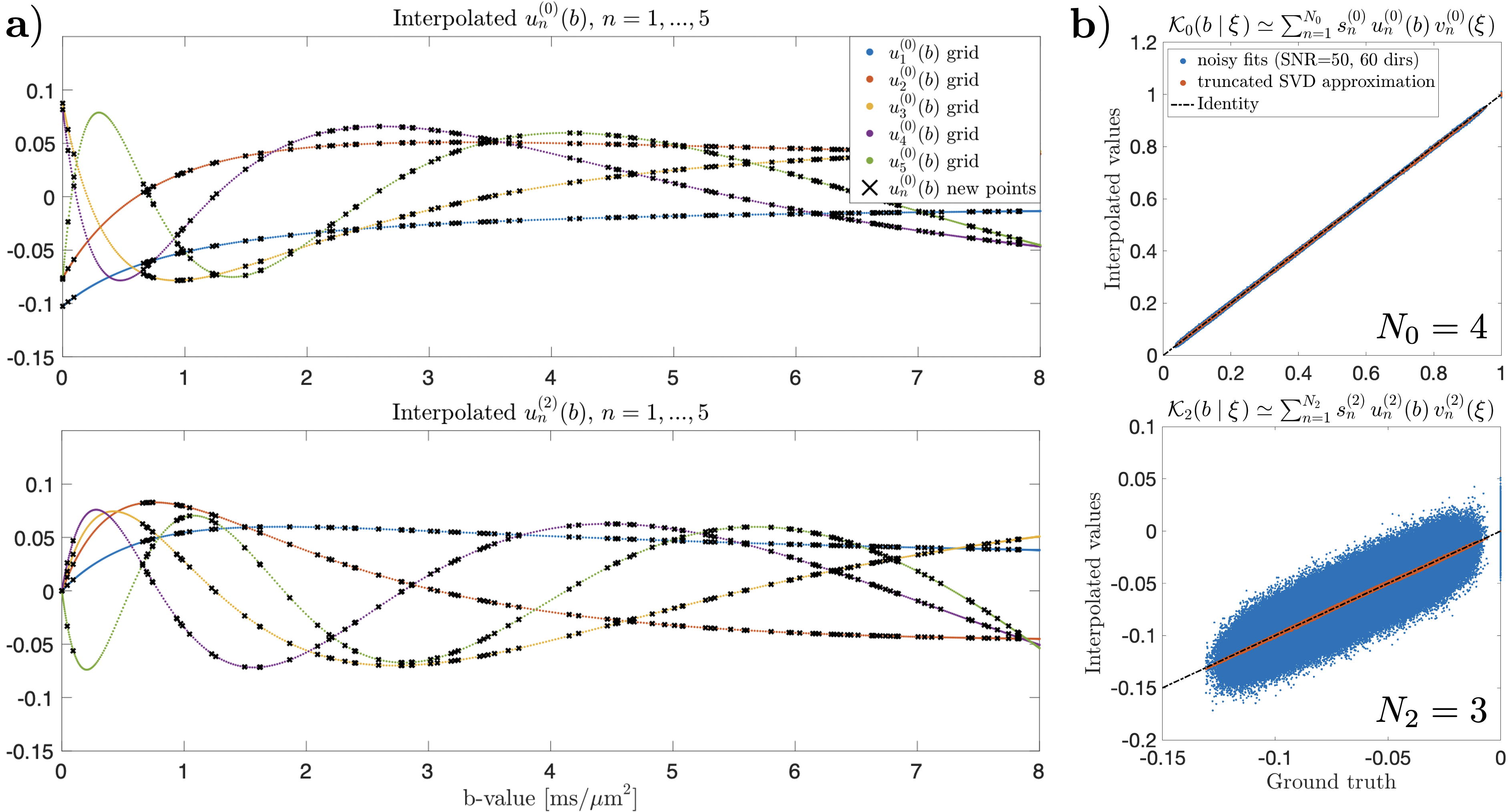}
	\caption[caption FIG 1]{(a) Chebyshev interpolation of basis functions $u_n^{(\ell)}(b)$ from SVD of a pre-computed library for the Standard Model, with 
	$\ell=0, \ 2$, and $N_0=N_2 = 5$ components for each $\ell$ in Eq.~(\ref{eq:KernelSVD}).   
	(b) Computation of rotational invariants $\mathcal{K}_\ell(b)$ for b-values outside the library based on interpolated $u_n^{(\ell)}(b)$. It can be seen that: (i) just a few SVD components already provide sufficient overall accuracy, (ii) the bias propagation from the interpolation of the basis functions to the rotational invariants is negligible for typical SNR conditions (SNR=50 at $b=0$, and 60 directions in a shell).}\label{fig:interpQuality}
\end{figure*}

\subsection{PIPE: tissue -- protocol factorization} 
\noindent
The  idea behind PIPE is to take the factorization (\ref{eq:SHfactorization}) a step further and fully decouple the kernel's dependence on tissue parameters $\xi$ and $p_{\ell m}$, and protocol parameters $b\to \{b, T_E, T_I, t, \dots \}$. 
This can be done numerically up to any desired accuracy by splitting the kernel's rotational invariants (\ref{eq:KrotInvs}) for a chosen tissue model (specified by a particular functional form of the kernel) into orthogonal functions using singular value decomposition (SVD):
\begin{equation}\label{eq:KernelSVD}
	\mathcal{K}_\ell(b  \mid \xi)\simeq\sum_{n=1}^{N_{\ell}}s_{n}^{(\ell)}\,u_{n}^{(\ell)}(b)\,v_{n}^{(\ell)}(\xi)
\end{equation}
up to arbitrary precision set by the number $N_\ell$. 
Substituting Eq.~(\ref{eq:KernelSVD}) into Eq.~(\ref{eq:SHfactorization}) and Eq.~(\ref{eq:SHfactorization2}), we obtain:
\begin{eqnarray}\nonumber
		S(b,\hat{\mathbf{g}}\mid\xi,p_{\ell m}) &\simeq& \sum_{n,\ell,m}\underbrace{u_{n}^{(\ell)}(b)\,Y_{\ell\,\!m}\left(\hat{\mathbf{g}}\right)}_{\alpha_{n\ell\,\!m}}\,\underbrace{s_{n}^{(\ell)}\,v_{n}^{(\ell)}(\xi)\,p_{\ell\,\!m}}_{\gamma_{n\ell\,\!m}}
		\\ \label{eq:alpha_gamma_eqn}
		&\equiv&   \sum_{n=1}^{N_\ell} \sum_{\ell =0}^{\ell_{\rm max}} \sum_{m=-\ell}^\ell \!\! \alpha_{n\ell\,\!m}(b,\hat{\mathbf{g}})\,\gamma_{n\ell\,\!m}(\xi,p_{\ell\,\!m}) \qquad
\end{eqnarray}
which allows us to write the dMRI signal as an expansion in the basis of data-driven functions, factorizing the protocol ($b,\hat{\mathbf{g}}$) and tissue ($\xi,p_{\ell m}$) dependencies. Indeed, the set of basis functions $\alpha_{n\ell\,\!m}(b,\hat{\mathbf{g}})$ depends purely on the protocol parameters, while  the set $\gamma_{n\ell\,\!m}(\xi,p_{\ell\,\!m})$ depends only on tissue (model) parameters. Including $s_{n}^{(\ell)}$ in the definition of $\gamma_{n\ell\,\!m}(\xi,p_{\ell\,\!m})$ ensures that the noise in the latter is approximately homoscedastic.

The factorization in Eq.~(\ref{eq:alpha_gamma_eqn}) decouples voxel-to-voxel variations in the protocol and tissue. Knowing the actual gradients (\ref{eq:Gactual}) everywhere in the FOV, we {\it interpolate} the local basis functions $\alpha_{n\ell\,\!m}\big(b(\r),\hat{\mathbf{g}}(\r)\big)$ (as described below) onto the unique set of diffusion encodings and directions for a  voxel at position $\r$, and {\it linearly} estimate tissue-dependent signal components 
\begin{equation}\label{eq:GammaVoxelwise}
	\hat{\gamma}_{n\ell\,\!m}(\r)=\alpha_{n\ell\,\!m}^\dagger(b,\hat{\mathbf{g}})\,\mathbf{S}(b,\hat{\mathbf{g}}) 
\end{equation}
using standard  Moore-Penrose pseudoinverse $\alpha_{\ell\,\!m n}^\dagger$ applied to the set of measurements $\mathbf{S}(b,\hat{\mathbf{g}}) = [ S(b_1,\hat{\mathbf{g}}_1) \hdots S(b_\mathrm{k},\hat{\mathbf{g}}_\mathrm{k})]^\mathrm{t}$ in a given voxel $\r$. 

Finally, we map the estimated $\hat\gamma_{n\ell\,\!m}(\r)$ onto the microstructural parameters of interest: 
\begin{equation}\label{ML}
\hat{\gamma}_{n\ell m}(\r) \ \rightarrow\  \{ \hat{\xi}(\r), \ \hat{p}_{\ell m}(\r) \} \,.
\end{equation}
Although  such mapping  is highly nonlinear, it is fully decoupled from the spatially-varying protocol. 
Hence, it can use the same trained regressor as $\r$ varies throughout the FOV.  
This allows us to learn the regression (\ref{ML}) {\it only once}, 
and apply it to $\hat{\gamma}_{n\ell m}$ from all brain voxels simultaneously in virtually no time. 
Practically, the SVD over the library of $\mathcal{K}_\ell(b\mid\,\!\xi)$ values used to compute $u_{n}^{(\ell)}(b)$ in Eq.~(\ref{eq:KernelSVD}) 
also provide us with the tissue  basis functions  $v_{n}^{(\ell)}(\xi)$.

From this point, we can proceed in two different ways. First, we combine  $v_{n}^{(\ell)}(\xi)$ and $s_{n}^{(\ell)}$  with randomly sampled fODF SH coefficients $p_{\ell m}$ to generate sets  
$(\gamma_{n\ell m} ; \ \  \{\xi, \ p_{\ell m}\})$ for the mapping (\ref{ML}), and use these as training data. Alternatively, we can form rotational invariants ${\gamma}_{n\ell}$ by factoring out $p_{\ell m}$ (akin to Refs.~\cite{REISERT2017,NOVIKOV2018}): 
The highest-SNR ones 
\begin{subequations} \label{gamma_1nell}
	\begin{equation}\label{gamma_1ell}
			{\gamma}_{1\ell} = \| \gamma_{1\ell m} \|_m = s_{1}^{(\ell)}\,v_{1}^{(\ell)}(\xi)\,p_{\ell}\,, \quad p_\ell = \| p_{\ell m}\|_m 
	\end{equation}
	for $n=1$, where $\|\dots\|_m$ is the 2-norm over $m=-\ell, \dots, \ell$,
	and $p_\ell$ are the fODF invariants; and the ones for $n>1$ by forming the ratios averaged over $m=-\ell, \dots, \ell$: 
	\begin{equation}\label{gamma_nell}
			{\gamma}_{n\ell} \Big |_{n>1}= 	\gamma_{1\ell}\left\langle \frac{\gamma_{n\ell m}}{\gamma_{1\ell m}} \right\rangle_m = s_{n}^{(\ell)}\,v_{n}^{(\ell)}(\xi)\,p_{\ell} \,. 
	\end{equation}
\end{subequations}
The invariants (\ref{gamma_1nell}) are analogous to those used in  Refs.~\cite{REISERT2017,NOVIKOV2018}, with the  dependence on discrete $b$ shells substituted by the SVD index $n$. \update{Like for (\ref{ML})}, we can learn the mapping 
\begin{equation}\label{ML_ell}
	\hat{\gamma}_{n\ell}(\r)\ \rightarrow \ \{ \hat{\xi}(\r), \ \hat{p}_{\ell}(\r) \} 
\end{equation}
from invariants (\ref{gamma_1nell})  to tissue parameters and fODF invariants.  \update{To solve either (\ref{ML}) or (\ref{ML_ell}), one needs to apply a sufficiently flexible regression. In this work, we employ the polynomial regression up to degree $W=3$}:
\begin{equation}
	\hat{\xi} = \sum_{j_1+j_2+\hdots + j_N \leq W} a_{j_1,j_2,\hdots,j_N} y_1^{j_1} \, y_2^{j_2}\,\hdots y_N^{j_N}\,,
\end{equation}
where $\hat{\xi}$ is the ML estimator, $\{y_i\}_{i=1}^N$ are the 
estimated $\gamma_{n\ell\,\!m}$, $W$ is the degree of the polynomial, and $a_{j_1,j_2,\hdots,j_N}$ are the regression coefficients computed during training.

\begin{figure*}[htbp]
	\centering
	\includegraphics[width=\linewidth]{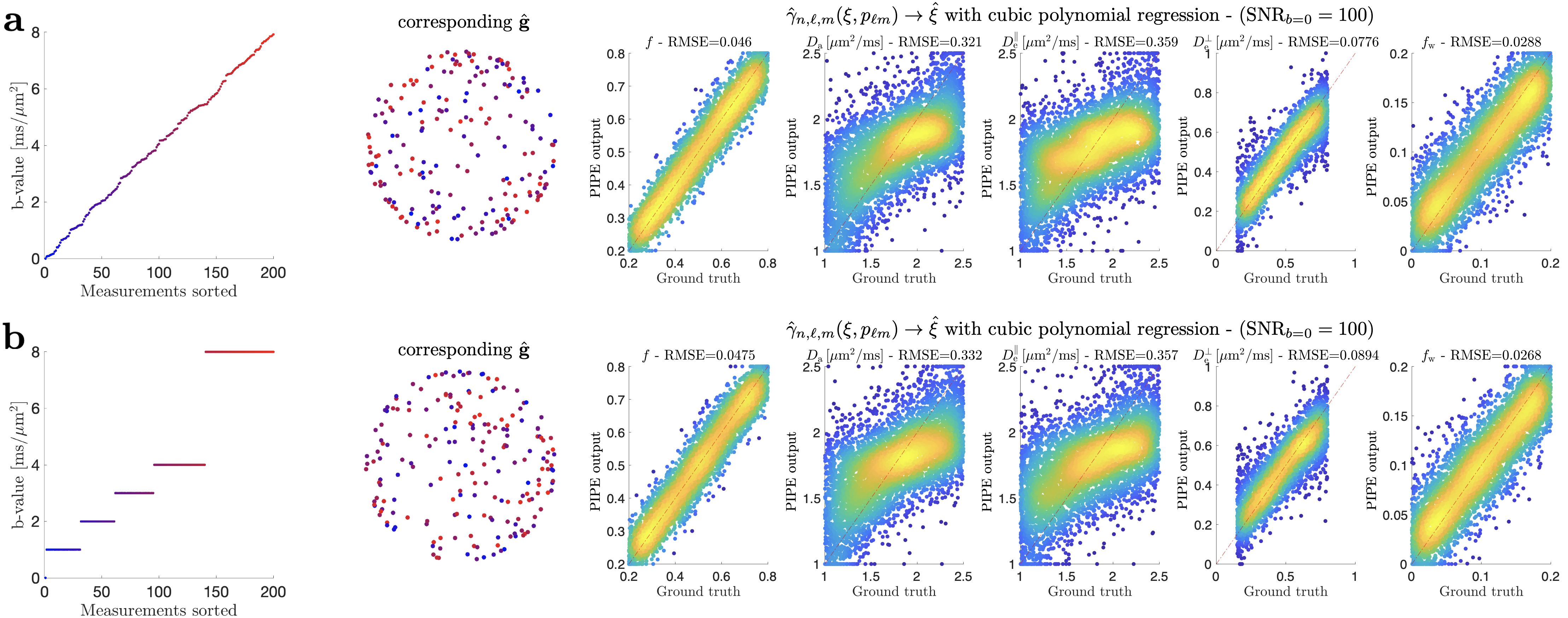}
	\caption[caption FIG 2]{Noise propagation \update{simulation} highlighting that PIPE can handle arbitrary protocols. Note that due to suboptimal experimental design (LTE data only), parallel diffusivities are highly dominated by the prior distribution used for training the machine learning estimator\cite{COELHO2021a}.}\label{fig:noisePropagation}
\end{figure*}

We observe that, for realistic SNR, mappings (\ref{ML}) and (\ref{ML_ell}) perform similarly, although (\ref{ML_ell}) is simpler due to the reduced dimensionality of ${\gamma}_{n\ell}$ compared to ${\gamma}_{n\ell m}$. Note that the mapping (\ref{ML_ell}) relies on having sufficient SNR to guarantee an unbiased $\hat{\gamma}_{1\ell}$ after taking the 2-norm over $m$ in Eq.~(\ref{gamma_1ell}). 
An alternative to Eqs.~(\ref{gamma_1nell}) could be to define ${\gamma}_{n\ell} =\| {\hat\gamma_{n\ell m}} \|_m = s_n^{(\ell)} v_n^{(\ell)}(\xi)p_\ell$ for all $n$, resembling the rotationally-invariant mapping\cite{NOVIKOV2018}. 
Practically, averaging the ratios ${\hat\gamma_{n\ell m}}/{\hat\gamma_{1\ell m}}$ over $m$ in Eq.~(\ref{gamma_nell}) helps mitigate the non-central-$\chi$ bias in the 2-norms $\| \hat\gamma_{n\ell m}\|_m$, as their SNR can become quite low with increasing $n$.

The interpolation of the protocol-dependent functions $\alpha_{n\ell\,\!m}(b,\hat{\mathbf{g}})$, Eq.~(\ref{eq:alpha_gamma_eqn}),  
for the specific directions $\hat{\mathbf{g}}$ 
is performed using spherical harmonics $Y_{\ell\,\!m}\left(\hat{\mathbf{g}}\right)$.   
Chebyshev interpolation\cite{TREFETHEN2019} is used for the continuum of $b$-values in the smooth functions $u_{n}^{(\ell)}(b)$, which we pre-compute at the Chebyshev polynomials' roots when generating the SVD factorization (\ref{eq:KernelSVD}) for a large matrix  containing a range of $\mathcal{K}_\ell(b\mid\,\!\xi)$ values (separately for each $\ell$),  for physically meaningful combinations of parameters $\xi$.

Figure \ref{fig:interpQuality} illustrates the SVD factorization (\ref{eq:KernelSVD}) for the Standard Model\cite{NOVIKOV2019} of diffusion in white matter, 
where 
\begin{equation}\label{eq:KernelLTE}
	\begin{aligned}
		\mathcal{K}(b,\g\cdot\n \mid \xi) 
		&= f \, e^{- bD_{{a}} (\g\cdot\n)^2} 		+ f_\text{w}\,e^{- b D_{\text{w}}}\\
		&+ (1-f - f_\text{w}) e^{- b(D_{e}^\parallel - D_{e}^\perp) (\g\cdot\n)^2  - bD_{e}^\perp  }
	\end{aligned}
\end{equation}
includes the intra-axonal compartment (a zero-radius ``stick" with a single tensor eigenvalue $D_a$ along the fascicle), the extra-axonal compartment (axially-symmetric tensor with axial and radial diffusivities $D_e^\parallel$ and $D_e^\perp$ oriented along the fascicle), and isotropic free water of diffusivity $D_\text{w}$ fixed at $3\,\mathrm{ms/\mu\,\!m^2}$, with the fractions 
$f+f_{e} + f_\mathrm{w} = 1$, such that $\xi$ is a set of the above diffusivities and fractions. 
Specifically,  we generated a library of $\mathcal{K}_\ell(b\mid \xi)$ up to  $\ell_\mathrm{max}=2$, 
consisting of 50,000 random sets of SM parameters (uniformly sampling $f\in[0.05,0.95],\,D_\mathrm{a}\in[1,3],\,D_\mathrm{e}^\|\in[1,3],\,D_\mathrm{e}^\perp\in[0.1,1.2],\,f_\mathrm{w}\in[0,1]$) and 1000 $b$-values sampled at the Chebyshev roots 
$\in[0,\, b_\mathrm{max}]$, with $b_\mathrm{max}=10\,\mathrm{ms/\mu\,\!m^2}$. 
This enabled accurate Chebyshev interpolation of $u_{n}^{(\ell)}(b)$ and subsequent approximation of $\mathcal{K}_\ell(b\mid\,\!\xi)$, see Fig.~\ref{fig:interpQuality}b (note that $\mathcal{K}_2<0$). 
This interpolation is performed for each voxel to obtain an exact protocol-dependent pseudoinverse $\alpha_{n\ell\,\!m}^\dagger(b,\hat{\mathbf{g}})$ entering Eq.~(\ref{eq:GammaVoxelwise}). 
Since both the interpolation and pseudoinversion are linear operations, this is not a computationally intensive step; it takes about the same time as the conventional DTI estimation.

\begin{figure*}[htbp]
	\centering
	\includegraphics[width=0.9\textwidth]{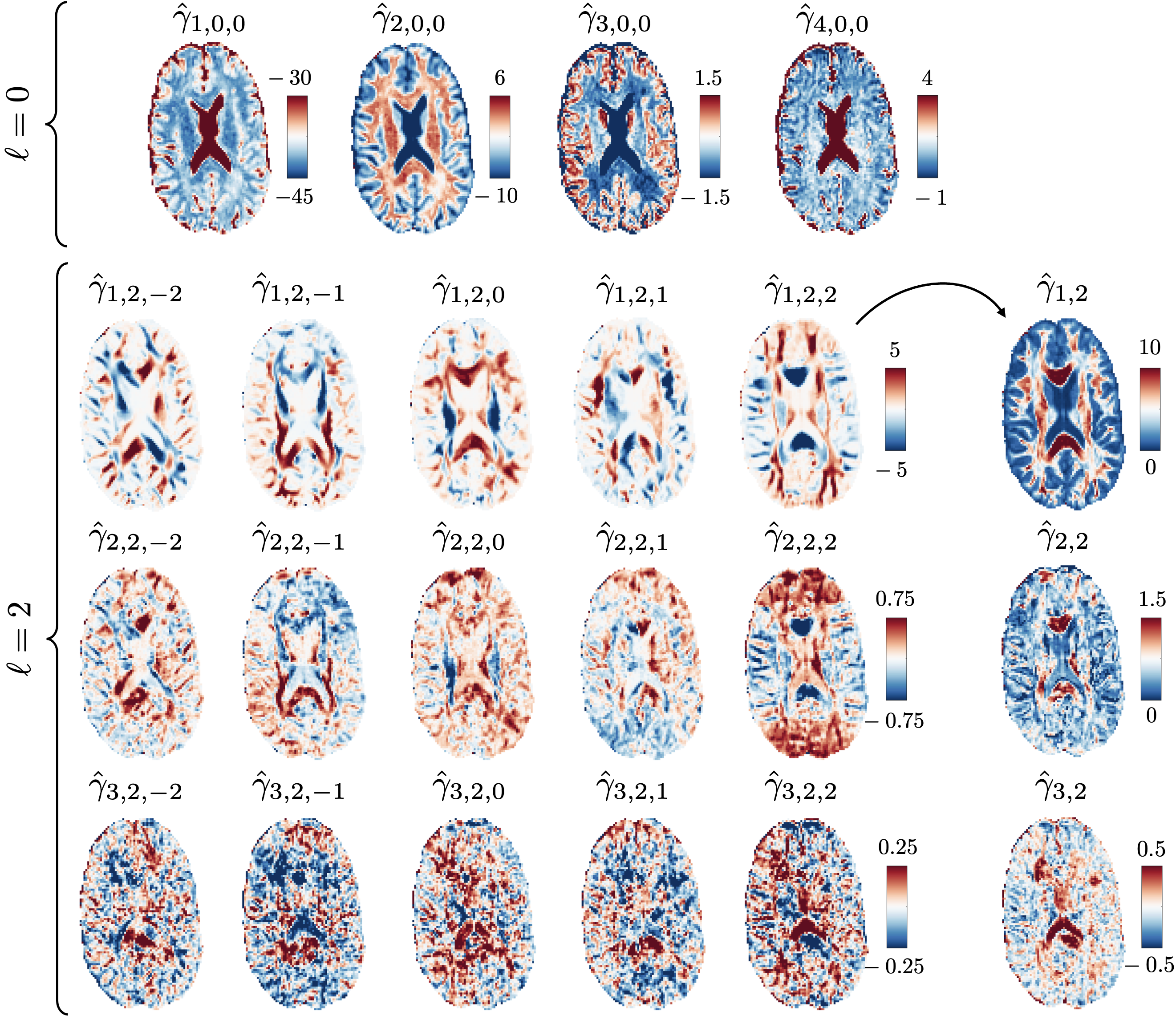}
	\caption[caption FIG 4]{Estimated $\hat{\gamma}_{n\ell\,\!m}$ maps for a healthy volunteer from non-shelled voxelwise protocols. Anatomical patterns similar to spherical harmonics $S_{\ell\,\!m}(b)$ are observed, with the difference that the $\hat{\gamma}_{n\ell\,\!m}$ do not depend on the protocol parameters because the combination of multiple $n$ captures the $b$-dependence. Larger $\ell$ values have fewer significant components $n$, and larger $n$ elements are noisier due to having smaller relative contributions (dictated by roughly exponentially decreasing $s_n^{(\ell)}$). $\gamma_{n\ell}$ denote the rotational invariants described in Eq.~(\ref{gamma_1ell})-(\ref{gamma_nell}).}\label{fig:gamma_hat_maps}
\end{figure*}

\subsection{MRI Experiments}
\subsubsection{Imaging}
\noindent
This study was performed under a local IRB-approved protocol. After providing informed consent, a 44 year-old male volunteer underwent MRI in a 3T-system (GE HealthCare, WI, USA) using an investigational MAGNUS head-only gradient insert\cite{FOO2020} and a 32-channel head coil (NOVA Medical, Wilmington, MA, USA). The system was powered by a 1 MVA gradient driver per gradient axis, operating at maximum gradient amplitude of $200\,\mathrm{mT}/\mathrm{m}$ and slew rate of $500\,\mathrm{T}/\mathrm{m}/\mathrm{s}$\cite{FOO2020}. A monopolar PGSE diffusion weighting sequence was used for acquiring shells at $(b[\mathrm{ms/\mu\,\!m^2}],\,N_\mathrm{dirs})=\{(1,25);(2,60);(8,50)\}$, with $\Delta=23\,\mathrm{ms}$, $\delta=12\,\mathrm{ms}$. Imaging parameters: voxel size$\,=2\times2\times2\,$mm$^3$, $\mathrm{TE}=46\,\mathrm{ms}$, $\mathrm{TR}=5\,\mathrm{s}$, echo-spacing$\,=\,424\,\mathrm{\mu\,\!s}$, in-plane acceleration$=2$, $\text{partial Fourier}\,=\,0.65$.

\subsubsection{Image reconstruction and processing}\label{ss:preproc}
\noindent
Diffusion-weighted complex MR images were reconstructed using GE Orchestra SDK tools including $k$-space filling approaches such as homodyne partial Fourier and ASSET. 
Denoising was applied in the complex domain\cite{VERAART2016,LEMBERSKIY2019}, therefore, reducing Rician bias significantly. 
Geometric distortions due to EPI were minimized due to the $500\,\mathrm{T}/\mathrm{m}/\mathrm{s}$ slew rate enabling a rapid readout, eliminating the need for distortion correction.

\begin{figure*}[htbp]
	\includegraphics[width=\linewidth]{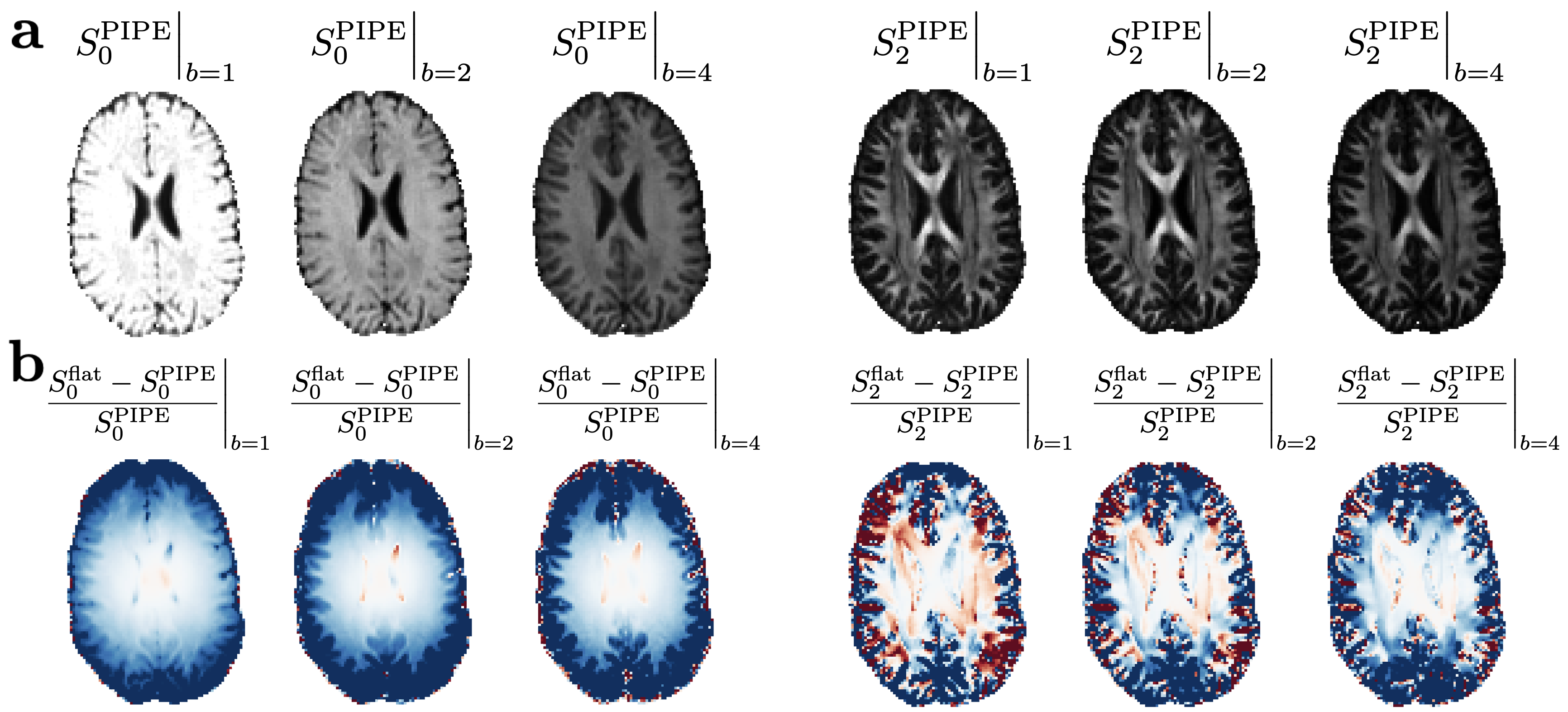}
	\caption[caption FIG 6]{a) Rotational invariants $S_0(b)$ and $S_2(b)$, normalized with $b=0$, computed from $\hat{\gamma}_{n\ell m}$ and resampling $\alpha_{n\ell m}(b,\hat{\mathbf{g}})$ at $b=1,\,2,\,4\, \mathrm{ms}/\mu\mathrm{m}^2$. b) Relative errors in $\%$ between accounting for gradient nonlinearities and considering a flat gradient field.}\label{fig:PIPE_vs_flat}
\end{figure*}

\update{In our experiments, subject motion was minimal and therefore no motion correction was applied. However, when motion is more pronounced, gradient nonlinearities interact with rigid-body head motion, modifying the alteration to the local $\B$-tensors differently for each diffusion-weighted image (DWI) \cite{RUDRAPATNA2021,GUO2021}. 
This effect can be corrected by applying the estimated rigid-body transformation of each DWI to the gradient coil tensor field, i.e. by evaluating 
$\L_{ij}$ at the original position 
$\r_0 = R^{-1}(\r-\mathbf{t})$, 
where $R$ and $\mathbf{t}$ are the rotation and translation from the acquired frame to the final frame. 
The motion-aware effective $\B$-tensor at voxel $\r$ is then given by $\L(\r_0)\B^\circ \L^t(\r_0)$, which then needs to be further rotated to match the updated image coordinate frame.
Incorporating such corrections is a natural extension of our framework and may be particularly important in studies involving pediatric or clinical populations where subject motion is unavoidable.}

\subsubsection{Model estimation and training data}\label{ss:paramEstimation}
\noindent
The linear estimation of $\hat{\gamma}_{n\ell\,\!m}$ from the set of diffusion measurements was done using a voxel-specific Moore-Penrose pseudoinverse, Eq.~(\ref{eq:GammaVoxelwise}). 
For the mapping (\ref{ML}), 
a cubic polynomial regression ($W=3$) was used for each model parameter in $\xi$, as it was deemed optimal for $25\leq\mathrm{SNR}\leq500$. Training data 
$\gamma_{n\ell m}= s_n^{(\ell)} v^{(\ell)}_n p_{\ell m}$ was used tor the regression to capture the factorization. 
Independent uniform priors were used for the kernel parameters  $f\sim\mathcal{U}(0.05,0.95),\,D_a\sim\mathcal{U}(0.5,3),\,D_\mathrm{e}^{\parallel}\sim\mathcal{U}(0.5,3),\,\,D_\mathrm{e}^{\perp}\sim\mathcal{U}(0.1,1.5),\,f_\mathrm{w}\sim\mathcal{U}(0,1)$, with $f+f_\mathrm{w}\leq 1$.  
The fODFs used in the training were random collections of $2$ fiber lobes with exponentially decaying $p_\ell$, as in Ref.\cite{CORONADO2024}: 
\begin{equation*}
\begin{aligned}
p_\ell &= C \lambda^{\ell},\quad p_2\sim \mathcal{U}(0.02,\,0.9),\,\lambda\sim \mathcal{U}(0.5,\,0.9),\\
p_{\ell m } &= w \, p_{\ell m }^{(1)}+ (1-w) \, p_{\ell m }^{(2)}, \quad w\sim \mathcal{U}(0,\,1)  .
\end{aligned}
\end{equation*}
The spherical harmonics coefficients of each lobe, $p_{\ell m }^{(1,2)}$ with $\ell\leq6$, were independently rotated in 3D using Wigner rotation matrices parametrized by ZYZ Euler angles $\alpha, \beta, \gamma$, \cite{Tinkham} with rotations uniformly sampled from the SO(3) group manifold $\mathbb{S}^3/\mathbb{Z}_2$ according to the invariant measure $\propto \sin \beta \,\d\alpha\d\beta\d\gamma$.  
Training took under 3 minutes for all kernel parameters. Computations were performed on a 3.7 GHz 6-Core i5 CPU with 32GB of RAM. 
All codes for PIPE were implemented in MATLAB (R2022a, MathWorks, Natick, Massachusetts). These are publicly available as part of the PIPE toolbox at \href{https://github.com/NYU-DiffusionMRI/PIPE}{https://github.com/NYU-DiffusionMRI/PIPE}.

\section{Results}\label{sec:Results}
\noindent
Field measurements from the head-only system used here show non-negligible gradient nonlinearities away from the isocenter, affecting peripheral parts of the brain. 
These can be observed by computing the re-scaling of the diffusion weighting $b(\r,\gz)/\bz$ as we traverse the FOV (this profile is specific to each nominal diffusion direction $\gz$), see Fig.~\ref{fig:gradientNonlinearities}a. With the irreducible decomposition in Eq.~(\ref{Ndecomp}), we can separate isotropic and anisotropic modifications to the nominal $b$-value due to nonlinearities, see Fig.~\ref{fig:gradientNonlinearities}b,c.

\begin{figure*}[htbp]
	\centering
	\includegraphics[width=0.7\linewidth]{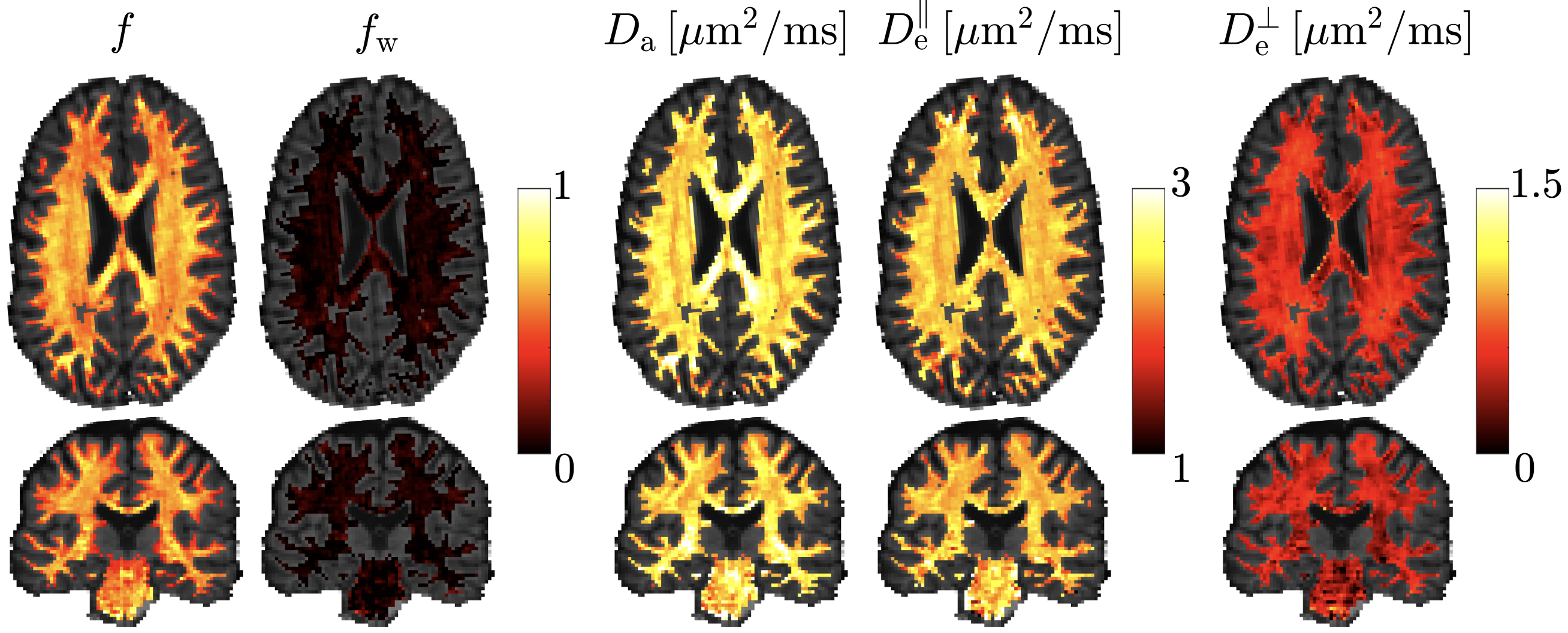}
	\caption[caption FIG 5]{Standard model microstructure maps ($\xi=[f,f_\mathrm{w},D_a,D_e^{||},D_e^{\perp}]$) for transversal and coronal slices. These maps were obtained by applying a fully connected neural network on the $\hat{\gamma}_{n\ell\,\!m}$ shown in Fig.~\ref{fig:gamma_hat_maps}. Note that this model only applies to white matter.}\label{fig:kernel_maps}
\end{figure*}

The accuracy of the Chebyshev interpolation of $u_n^{(\ell)}(b)$ library values is shown in Fig.~\ref{fig:interpQuality}a. Additionally, Fig.~\ref{fig:interpQuality}b indicates that $N_0\!=\!4$ and $N_2\!=\!3$ basis functions suffice for accurate factorization of the Standard Model kernel in white matter. Errors due to SVD truncation and interpolation of  $u_n^{(\ell)}(b)$ are below typical SNR levels and, therefore, can be considered negligible. This is possible due to the  exponentially fast decrease of $s_n^{(\ell)}$ with $n$.

Noise propagation experiments emphasize the flexibility of our proposed approach to work on shelled and non-shelled data, see Fig.~\ref{fig:noisePropagation}. This showcases that PIPE can account for arbitrarily large gradient nonlinearities, applying even to cases where the nominal shelled structure of data is lost due to nonlinearities. In this work we have focused on the Standard Model of diffusion in white matter (\ref{eq:KernelLTE}) as an illustration, but the PIPE framework can be applied to any spherical convolution-based approach\cite{TOURNIER2004,ANDERSON2005,TOURNIER2007,JEURISSEN2014,JESPERSEN2007,FIEREMANS2011,ZHANG2012,KADEN2016,JENSEN2016,REISERT2017,NOVIKOV2018,VERAART2017,PALOMBO2020,JELESCU2022,OLESEN2022} irrespective of the functional dependence of the kernel on diffusion weighting parameters.

\textit{In vivo} full brain maps of the linearly estimated $\hat{\gamma}_{n\ell\,\!m}$ parameters are shown in Fig.~ \ref{fig:gamma_hat_maps} for a healthy volunteer. These parameters approximate the nonlinear $b$-value dependence of $\mathcal{K}_0(b)$ and $\mathcal{K}_2(b)$ rotational invariants of the Standard Model kernel (\ref{eq:KernelLTE}). The resemblance between $\hat{\gamma}_{n,2,m}$ and maps of spherical harmonics is due to the fODF $p_{2m}$ coefficients being part of the former. Neglecting the spatial variation of the diffusion encoding causes a bias dependent on the distance of the voxel to the bore's isocenter. 

Figure \ref{fig:PIPE_vs_flat}a shows signal rotational invariants computed by resampling $\alpha_{n\ell m}(b,\hat{\mathbf{g}})$ at $b=1,\,2,\,4\, \mathrm{ms}/\mu\mathrm{m}^2$ uniformly over the unit sphere and applying Eq.~(\ref{eq:alpha_gamma_eqn}) to get $S(b,\hat{\mathbf{g}})$ and subsequently $S_\ell(b)$. The effect of accounting for gradient nonlinearities is minor in the center of the brain because it coincided with the bore's isocenter, while its significance gradually increase up to values of the order $10\%$. 
Corresponding parameter biases will likely be greater due to a highly nonlinear parameter estimation for typical brain microstructure models. 
Figure \ref{fig:kernel_maps} shows SM parametric maps obtained after applying a polynomial regression to $\hat{\gamma}_{n\ell\,\!m}$.

\section{Discussion}\label{sec:Discussion}
\noindent
Gradient nonlinearities are ubiquitous in MRI systems, a natural consequence of the Maxwell's equations. A design variable is the linearity region, which determines the size of the organs that can be scanned (head-only vs whole-body scanners). 
Smaller linearity regions for MRI gradient coils offer advantages for achieving higher gradient performance, particularly in terms of gradient strength and slew rate. This has motivated recent hardware advances in head-only\cite{HUANG2021,RAMOS2025} or insertable gradient coils\cite{FOO2020,FEINBERG2023}, optimized for brain imaging. These provide access to the frontier of tissue microstructure imaging through the lens of diffusion; however, this comes at the cost of FOV spatial variations in diffusion weightings. 

This work addresses the practical challenge of efficiently handling variations in diffusion weightings due to gradient nonlinearities. 
\update{Previous work exploited the linearity of the DTI log-signal to correct diffusion tensors post hoc\cite{BAMMER2003}. Such corrections, however, are not applicable to more general nonlinear models. In these cases, $\B$-tensors must be corrected in a voxelwise fashion before the estimation process, making data driven estimators extremely inefficient.} 
The proposed PIPE framework enables fast parameter estimation of a wide class of biophysical models (spherical convolution models) from dMRI data where each voxel is acquired with a different diffusion weightings and directions.

The key ingredient behind PIPE is the factorization of the signal dependence on experimental and tissue parameters. 
This is achieved through SVD and a projection into spherical harmonics basis. Therefore, allowing us to split the estimation process into two parts: (i) a linear step that handles the spatial variations of experimental parameters while remaining independent of tissue parameters; (ii) a nonlinear step that converts SVD coefficients into model parameters. 

Training compute time for step (ii) was fast ($\sim1$-minute) because PIPE requires to be trained only once, irrespective of the magnitude of gradient nonlinearities. Naively retraining conventional fixed-protocol ML estimators for each voxels would take a prohibitive amount of compute ($\sim 100$ days considering a full brain at 2mm resolution has in the order of $10^5$ voxels in the brain). 
Furthermore, accuracy and memory consumption are significantly improved with respect to computing a brute-force regression where the full experimental and model parameter space are sampled jointly.

The PIPE framework can be tailored to any model based on a spherical convolution of a fiber response with an fODF\cite{TOURNIER2004,ANDERSON2005,TOURNIER2007,JEURISSEN2014,JESPERSEN2007,FIEREMANS2011,ZHANG2012,KADEN2016,JENSEN2016,REISERT2017,NOVIKOV2018,VERAART2017,PALOMBO2020,JELESCU2022,OLESEN2022}, thereby allowing its application in white and gray matter. Additionally, PIPE is readily extendable for simultaneously varying diffusion times, TE, etc. PIPE does not require the data to be sampled in shells. Furthermore, it provides optimal data resampling for those algorithms that do need shells, e.g.,  Refs.~\cite{TOURNIER2007,JEURISSEN2014}. This feature is shared with previously proposed approaches focused on data resampling for artifact correction \cite{CHRISTIAENS2019}. However, the main advantage of PIPE is that it is not constrained to shells and that it allows for spatially varying protocols throughout the FOV.

We have tested the feasibility of our method on simulations and \textit{in vivo} brain dMRI data acquired with a head-only gradient insert with non-negligible gradient nonlinearities. Results showed no trace of gradient nonlinearities in output maps, see Fig~\ref{fig:kernel_maps}. 
Our results will stimulate the development of high-performance gradient hardware where nonlinearities are allowed to be large by design.

The proposed estimator, like any other data-driven algorithm, is influenced by the nature of its training data. The dependence on the training set is tied to the quality of acquisitions (sampling of the $q$-space and the SNR). In scenarios where comprehensive protocols are employed, the algorithm's performance is less susceptible to variations in training data\cite{COELHO2021a}. Conversely, in cases with limited protocols containing less information, the performance is more biased towards the mean of the training set\cite{COELHO2021a}. Here, we used independent uniform distributions for training to minimize spurious parameter correlations.

The impact of gradient nonlinearities on multidimensional diffusion encodings to date has focused on their effect on Maxwell compensation for asymmetric gradient waveforms\cite{SZCZEPANKIEWICZ2020}. In this work we provide the first in-depth analysis of the impact of gradient nonlinearities on arbitrary multidimensional diffusion encodings\update{, extending previous work on LTE \cite{BAMMER2003}. Our} 
result provides a parsimonious way to separate isotropic and anisotropic contributions of gradient nonlinearities to the $b$-value and $\B$-tensor shape.  \update{Both isotropic and anisotropic components affect the $b$-value, whereas only anisotropic components alter the $\B$-tensor shape, see Appendix \ref{AppBactual}.}  
The advantage of acquiring LTE diffusion encodings  is that while gradient nonlinearities change the actual intensity and direction of the diffusion weighting, they do not affect the $\B$-tensor shape. One should be cautious when acquiring non-LTE data on systems with strong gradient nonlinearities as these will introduce deformations in $\B$-tensor, removing any pre-set symmetries. This would require a re-definition of the convolution on the SO(3) group manifold rather than on a 2-dimensional sphere, which is beyond the scope of this work. Approaches like Ref. \cite{SZCZEPANKIEWICZ2019} could be used to design encodings that remain maximally symmetric for specific nonlinearities.

\section{Conclusions}\label{sec:Conclusion}
\noindent
We proposed a two-step machine learning PIPE parameter estimation framework that enables fast parameter estimation of convolution-based diffusion MRI models from data where each voxel is acquired with a different protocol. This method allows for the straightforward application of a large class of biophysical tissue models to data acquired with arbitrarily large gradient nonlinearities. Furthermore, the data is not constrained to be acquired in any fashion, e.g., in shells, and there are no limits to the gradient nonlinearities as long as the protocol is well defined for each voxel. This method is readily extendable for simultaneously varying diffusion times, echo or inversion times.

\section*{Acknowledgments}
This work has been supported by NIH under NINDS award R01 NS088040 and NIBIB awards R01 EB027075, P41 EB017183, K99 EB036080, and NIH OD and NIDCR award DP5OD031854. The authors are grateful to Sune Jespersen for fruitful discussions, and to Eric Fiveland, Maggie Fung, and Chitresh Bhushan for their help with the MRI experiments.

All processing codes for the protocol independent parameter estimation (PIPE toolbox) are available at \href{https://github.com/NYU-DiffusionMRI}{https://github.com/NYU-DiffusionMRI}.



\subsection*{Conflict of interest}
GL, EF, DSN, and NYU School of Medicine are stock holders of MicSi, Inc. --- post-processing tools for advanced MRI methods. 
SC, EF, and DSN are co-inventors in technology related to this research; a PCT patent application has been filed.  
AZ, NA, and TKFF are employees of GE HealthCare. HHL is an equity holder for NVIDIA, Corp.

\appendix
\section{: How an arbitrary $\B$-tensor is modified by gradient nonlinearities}\label{AppBactual}
\noindent
The most general  (nominal) $\B$-tensor, a symmetric $3\times 3$ matrix with 6 independent parameters, admits the irreducible decomposition (in the notations of Ref.~\cite{COELHO2025}):
\begin{equation}
	\begin{aligned} \label{Basym}
		\Bz_{ij} =   \frac{\bz}3\,\delta_{ij}+ \sum_{m=-2}^2 \Bz^{2m} \Y^{2m}_{ij} \,.
	\end{aligned}
\end{equation}
Here 
the first term 
defines the isotropic part, and the remaining five parameters $\Bz^{2m}$, coupled to symmetric trace-free tensors $\Y^{2m}_{ij}$, characterize the anisotropy of encoding while not contributing to the overall diffusion weighting $\bz=\tr \Bz$.  
Substituting Eq.~(\ref{Basym}) into Eq.~(\ref{Bactual}), the actual $\B$-tensor for this general case becomes
\begin{equation} \label{Basym_L}
\B = \frac{\bz}3 \, \L\L^t + \sum_{m=-2}^2 \Bz^{2m}\, \L\, \Y^{2m}\, \L^t \,.
\end{equation}
We can see that the tensor $\N=\L^t \, \L$ introduced in the main text is not enough to describe the deformation of a  generic $\Bz$. 

Let us find the $\ell=0$ and $\ell=2$ spherical tensor components of the symmetric tensor (\ref{Basym_L}). 
The $\ell=0$ isotropic part $\frac{b}3 = \frac13 \tr \B$, using cyclic permutation under the trace, can be cast via the components of the $\N$ tensor: 
\begin{eqnarray}  \nonumber
b = \tr \B &=& \bz \N_0 + \sum_{m=-2}^2 \Bz^{2m} \tr \lp \Y^{2m}\N \rp \\
&=& \bz \N_0 + \tfrac32 \sum_{m=-2}^2 \Bz^{2m} \N^{2m*}  \,, 
\label{b_L}
\end{eqnarray}
where we used the fact that $\N$ is real-valued, and the orthogonality\cite{COELHO2025} $\Y^{2m*}_{ij} \Y^{2m'}_{ij} = \tfrac32\delta_{mm'}$.  
The geometric meaning of Eq.~(\ref{b_L}) is as follows. The  $\ell=0$ component of (\ref{Basym_L}) comes from two sources: the $\ell=0$ component of $\N$, and the $\ell=0$ component of the ``addition of angular momenta", $\vec{2}\otimes\vec{2}\to\vec{0}$, of the $\ell=2$ components of $\Bz$ and $\N$.
Indeed, using $\N^{2m*}=(-1)^m \N^{2,-m}$, the sum in  Eq.~(\ref{b_L}) can be understood as\cite{COELHO2025}
\begin{eqnarray}\nonumber
&& \tfrac15  \sum_{m=-2}^2 \Bz^{2m} \N^{2m*} = \tfrac15 \sum_m (-1)^m \Bz^{2m} \N^{2,-m}   \\
&=&  \nonumber
 \braket{ 2 0 2 0 | 0 0}  \sum_{m,m'} \braket{2 m 2 m' | 0 0} \Bz^{2m} \N^{2m'}   \\
&=&  \lb \Bz^{(2)}\otimes \N^{(2)} \rb_{\ell=0} \,,   
\end{eqnarray}
where the Clebsch-Gordan coefficients $\braket{ 2 0 2 0 | 0 0}=1/\sqrt{5}$ and $\braket{2 m 2 m' | 0 0} = (-1)^m \delta_{m,-m'}/\sqrt{5}$. 

The $\ell=2$ components are found via $\B^{2m}=\tfrac23 \Y_{ij}^{2m*}\,\B_{ij}$:
\begin{equation} \label{B2m_L}
\B^{2m} \!=\! \tfrac29\bz \tr \lp \Y^{2m*}\L\L^t\rp + \tfrac23 \sum_{m'} \Bz^{2m'} \tr \lp \Y^{2m*} \L \,\Y^{2m'} \L^t\rp 
\end{equation}
They stem both from the $\ell=2$ anisotropy of symmetric matrix $\L\L^t$, 
and from the $\ell=2$ part of the object $\L\, \Bz^{(2)}\, \L^t$, where tensors $\L$ and $\L^t$ have the same $\ell=0,2$ components, 
and the $\ell=1$ components $a_i$ and $-a_i$, respectively, where the pseudo-vector $a_i = \epsilon_{ijk}L_{jk}$. 

From the  result (\ref{B2m_L}) we observe that the presence of generic matrices $\L$ and $\L^t\neq \L$, which do not possess any symmetry and have all their $1+3+5=9$ components corresponding to the irreducible representations with $\ell=0, \ 1, \ 2$, removes any special symmetry of the nominal $\Bz$ tensor (\ref{Basym}). 
For example, even if the original $\Bz =\tfrac{\bz}3 \delta_{ij}$ were isotropic, $\Bz^{2m}\equiv 0$, the anisotropy of the symmetric tensor $\L\L^t$ (nonzero $m=-2\dots 2$ components  for $\ell=2$) means that the actual $\B$-tensor will be of a generic form, without any particular symmetries. The same applies to the widely used  axially symmetric family\cite{TOPGAARD2017,WESTIN2016} 
\begin{equation}\label{eq:AxSymB}
	\Bz_{ij}=\bz\,\Big( \,\betaz \, \gz\otimes\gz + \tfrac{1-\betaz}{3} \, \mathsf{I} \Big) 
\end{equation}
parametrized by the unit vector $\gz$ along the symmetry axis and by the shape parameter $\betaz$. In the basis where $\gz = \hat{\bf z}$, 
only ${\Bz}^{20}$ will be nonzero, out of five ${\Bz}^{2m}$. However, for a generic $\L$, Eq.~(\ref{B2m_L}) will have all the actual $\B^{2m}$ components $\propto \tr \lp \Y^{2m*} \L Y^{20} \L^t\rp $ present. 
This has important implications, since sampling with generic $\B$-tensors takes us away from the convolution on a 2-sphere $\mathbb{S}^2$ onto the 3-dimensional SO(3) group manifold. 

\bibliography{Coelho_bibliography_2022_04_19.bib}

\end{document}